\documentclass[11pt]{article}
\usepackage{fancyhdr}
\usepackage{isomath}
\usepackage{amsmath}
\usepackage{amsbsy}
\usepackage{amssymb}
\usepackage{amscd}
\usepackage{amsfonts}
\usepackage{graphicx,color}
\usepackage{verbatim}
\usepackage{euscript}
\usepackage{alltt}
\usepackage{stmaryrd}
\usepackage{subfigure}
\usepackage[utf8]{inputenc}
\usepackage[T1]{fontenc}

\newcommand{\divv}{\mbox{div} \,}
\newcommand{\curl}{\mbox{curl} \,}

\newcommand{\Va}{V^{\alpha}}
\newcommand{\tr}[1]{\mbox{tr}(#1)}
\newcommand{\dpsi}[1]{\partial_{#1} \psi}
\newtheorem{remark}{Remark}
\newcommand{\C}{\mathbb C}
\newcommand{\R}{\mathbb R}
\newcommand{\calC}{\mathcal C}

\usepackage[margin=1in]{geometry}

\date{}
\begin{document}
\title{Coupled Dislocations and Fracture dynamics at finite deformation:\\ model derivation, and physical questions}
\author{Amit Acharya\thanks{Dept. of Civil \& Environmental Engineering, and Center for Nonlinear Analysis, Carnegie Mellon University, Pittsburgh,USA, email: acharyaamit@cmu.edu}} 
\maketitle

$\quad $ \textit{This paper is dedicated to Professor Nasr Ghoniem on the occasion of his retirement.}

\begin{abstract}
\noindent A continuum mechanical model of coupled dislocation based plasticity and fracture at finite deformation is proposed. Motivating questions and target applications of the model are sketched.

\end{abstract}

\section{Introduction}
Based on experience and insights gathered from the partial differential equation (PDE) based modeling of dislocation dynamics in \cite{zhang2015single, arora2020unification, arora2020finite, garg2015study} and fracture \cite{acharya2018fracture, acharya2020possible, morin2021analysis}, a coupled model of fracture and dislocation based plasticity at finite deformation is explored. Even though plasticity, whether fundamentally rooted in the mechanics of dislocations or in the phenomenology of slip, and fracture are much studied subjects, e.g.~\cite[and the literature reviews in the papers mentioned above]{freund1998dynamic, hutchinson_fracture, hirth_lothe, bulatov2006computer, asaro1983micromechanics, havner1992finite}, to our knowledge, a full-blown continuum PDE model for their coupled mechanics does not exist and can be useful in the understanding of the deformation, flow, and fracture of solids (e.g., metals or glaciers), and the mutual interactions of these phenomena as, e.g., addressed in the seminal work \cite{rice1974ductile}. While a full thermodynamically consistent model is presented, it is recognized that this is merely a beginning that sets the stage for future computation and analysis of a simply stated, but intricate, nonlinear model which is expected to have some bearing on its target applications.

An outline of the paper is as follows: in Sec.~\ref{sec:mech} the mechanical equations of the model are presented. In Sec.~\ref{sec:thermo} a possible set of thermodynamically consistent constitutive equations are proposed. In Sec.~\ref{sec:motiv} some target problems motivating the development of the theory are sketched. It is understood that most of the questions posed are beyond the reach of rigorous methods of PDE analysis, but it is felt that demonstrating a dynamical theoretical setup where such questions can at least begin to be clearly posed and, consequently, at least be approached with finite-dimensional methods of approximation and rigorous mathematical guidance, even if short of the `proven-theorem' variety, can be helpful in advancing the science of deformation, flow, and fracture of solids.

A few words on notation: tensor components (when invoked) are written with respect to the basis of a fixed Rectangular Cartesian coordinate system. All spatial differential operators are w.r.t. position on the current configuration. A superposed dot represents a material time derivative. $X$ will be the alternating tensor and the $curl$ operator acting on tensor fields may simply be thought of as row-wise curls of the corresponding matrix field of components.

\section{Governing Equations: Mechanics}\label{sec:mech}
Based on the detailed kinematic motivations presented in \cite{acharya2011microcanonical,acharya2018fracture,acharya2020possible}, the governing equations of the model are given by
\begin{subequations}
\label{eq:gov_eq}
\begin{align}
\dot{\rho} + \rho \, \divv v & = 0 \qquad \qquad & \mbox{(Balance of mass)}\\
\rho \dot{v} & = \divv T + \rho f \qquad \qquad & \mbox{(Balance of linear momentum)}\label{eq: blm}\\
\dot{W} + WL &= - \curl W \times \Va + L^p \qquad \qquad & \mbox{(Evolution of inverse elastic distortion)}\label{eq:W_evol}\\
\dot{c} + L^T c & = - \curl c \times V^t, \qquad \qquad & \mbox{(Evolution of the crack field)}\label{eq:c_evol}
\end{align}
\end{subequations}
where $T = T^T$ ensures balance of angular momentum, and
\[
L := \nabla v
\]
is the velocity gradient. In the above, $T$ is the Cauchy stress, $\rho$ is the mass density, $v$ is the material velocity, $f$ is the prescribed body force density, $W$ is the inverse elastic distortion (a 2-point tensor field), $\Va$ is the dislocation velocity (vector field), $L^p$ is a meso-macroscale construct not used in the fundamental microscale theory, the plastic distortion rate of dislocations (tensor field) that are `averaged out' in terms of their charge (the meaning of this can be made precise in terms of microscopic quantities), $c$ is the crack (vector field), and $V^t$ is the crack-tip velocity (vector field). The magnitude of the crack vector field encapsulates the degree of damage at a material point, while its orientation reflects that of the crack face normal at that point. An independent vector-valued field representing the crack-face normal as a fundamental kinematic ingredient in a PDE model of fracture was introduced in \cite{acharya2018fracture,steinke2019phase},  and is beginning to be used \cite{morin2021analysis, hakimzadeh2022phase,steinke2022energetically}. The dislocation and crack-tip velocity fields are relative velocities of the motion of the dislocation density field $\alpha$ and the crack-tip field $t$, respectively, w.r.t.~the material velocity. Defining the \emph{dislocation} and \emph{crack-tip} line density fields
\begin{equation}\label{eq:line_den}
- \curl W =: \alpha; \qquad - \curl c =: t
\end{equation}
\eqref{eq:W_evol} and \eqref{eq:c_evol} imply
\begin{subequations}
\label{eq:conservlaw}
\begin{align}
 \mathring{\alpha}  &:= \dot{\alpha} + \tr{L} \alpha - \alpha L^T   = -\curl \left( \alpha \times \Va + L^p \right) \label{eq:a_evol}\\
 \mathring{t} & := \dot{t} + \tr{L} t - Lt  =  - \curl \left(t \times V^t \right) \label{eq:t_evol}
\end{align}
\end{subequations}
Physically, (\ref{eq:W_evol}, \ref{eq:c_evol}) are motivated from the conservation of topological charge statements (\ref{eq:a_evol}, \ref{eq:t_evol}) on assuming that a `free' gradient that arises in the process in each case vanishes.

In the above, $W$ and $c$ are considered to be dimensionless physical quantities.
\section{Guidance for constitutive assumptions from the Second Law of Thermodynamics}\label{sec:thermo}
We consider the free-energy density per unit mass 
\begin{equation}\label{eq:free_energy}
\psi = \psi(W, \alpha, c, t, \rho)
\end{equation}
and require that the power supplied by external agents be greater than or equal to  the rate of change of the sum of the free energy and kinetic energy of the body:
\begin{equation} \label{eq:sec_law}
\begin{split}
&\int_{\partial \calC} (Tn)\cdot v \, da + \int_{\mathcal{C}} \rho f \cdot v \, dv \geq \frac{d}{dt} \left( \int_{\mathcal{C}}  \rho \psi \, dv + \int_{\mathcal{C}} \frac{1}{2} \rho |v|^2 \, dv\right) \\
&\Rightarrow \int_{\mathcal{C}} T:L\, dv - \int_{\mathcal{C}} \rho \dot{\psi} \, dv \geq 0
\end{split}
\end{equation}
for any process in which the mechanical equations hold, where $\calC$ is the (time-varying) current configuration of the body; and this is ensured by choosing constitutive assumptions for $T, \Va, V^c, L^p$ that guarantee \eqref{eq:sec_law}.

Now,
\begin{equation*}
\begin{split}
\rho \dot{\psi} =  & \, \rho\left( \dpsi{W} : \dot{W} + \dpsi{\alpha} :\dot{\alpha} + \dpsi{c} \cdot \dot{c} + \dpsi{t}\cdot \dot{t} + \dpsi{\rho} \dot{\rho} \right)\\
= & \, \rho \, \dpsi{W} : \left( - WL + \alpha \times \Va + L^p \right)\\
& + \rho \, \dpsi{\alpha} : \left(- \alpha (L:I) + \alpha L^T - \curl(\alpha \times \Va + L^p ) \right)\\
& + \rho \, \dpsi{c} \cdot ( - L^T c + t \times V^t )\\
& + \rho \, \dpsi{t} \cdot (- t(L:I) + Lt - \curl(t \times V^t) )  \\
& + \rho \dpsi{\rho} (- \rho (L:I) ).
\end{split}
\end{equation*}
so that \eqref{eq:sec_law} can be expressed as 
\begin{equation}\label{eq:sec_law_exp}
\begin{split}
& \int_{\mathcal{C}} \left[ T + {\color{blue} \rho\left\{W^T \dpsi{W} + (\dpsi{\alpha}:\alpha) I + c \otimes \dpsi{c} - \dpsi{\alpha}^T \alpha + (\dpsi{t}\cdot t) I - \dpsi{t} \otimes t + \rho \dpsi{\rho} I\right\} } \right]:L \,dv\\
&+ \int_{\mathcal{C}} \left[ - \rho \,\dpsi{W} + \curl(\rho \, \dpsi{\alpha}) \right] : L^p \, dv\\
& + \int_{\mathcal{C}} \left[ X \left\{- \rho \, \dpsi{W} + \curl (\rho \, \dpsi{\alpha}) \right\}^T \alpha  \right] \cdot{\Va} \, dv\\
& + \int_{\mathcal{C}} \left[ \left\{ - \rho \dpsi{c} + \curl (\rho \dpsi{t}) \right\} \times t \,\right] \cdot V^t \, dv \\
& + \int_{\partial \mathcal{C}} (\rho \dpsi{\alpha} \times n): L^p \, da +  \int_{\partial \mathcal{C}} \left[ \left( \rho \dpsi{\alpha}  \times n \right)^T \alpha\right]  \cdot V^{\alpha} \, da + \int_{\partial \mathcal{C}} [(\rho \dpsi{t} \times n) \times t] \cdot V^t \, da \\
& \quad \geq 0.
\end{split}
\end{equation}
Since the response function for $\psi$ is invariant under superposed rigid body motions, it can be shown (see Appendix) that the term highlighted in blue in \eqref{eq:sec_law_exp} is symmetric. Since the stress is symmetric due to balance of angular momentum, this also implies that the spin ($L_{skw}$) does not appear in the dissipation for the model making the latter invariant under superposed rigid motions. This is an important consistency check on the kinematic structure of the model.

Viewing $L^p, V^\alpha$, and $V^t$ (in the bulk and at the boundary) as the sole dissipative mechanisms of the model, one recovers the stress-relation of the model from the consideration of energetically reversible, purely elastic processes:
\begin{equation}\label{eq:stress_reln}
T = - {\rho\left\{W^T \dpsi{W} + (\dpsi{\alpha}:\alpha) I + c \otimes \dpsi{c} - \dpsi{\alpha}^T \alpha + (\dpsi{t}\cdot t) I - \dpsi{t} \otimes t + \rho \dpsi{\rho} I\right\} }_{sym}.
\end{equation}
Finally, a sufficient condition for non-negative dissipation is obtained by choosing the constitutive assumptions for the dissipative mechanisms to be in the `direction' of their respective `driving forces,' as exposed in \eqref{eq:sec_law_exp}, in the bulk and at the boundary.
\subsection{A specific set of thermodynamically consistent constitutive assumptions}
Define
\[
F := W^{-1}, \qquad \qquad E := \frac{1}{2} \left( F^TF - I \right), \qquad \qquad r:= F^T c,
\]
(and we caution that $F$ is \textbf{\textit{not, in general, the gradient of a deformation w.r.t.a fixed global reference configuration}}). With $\mathbb{I}$ the fourth-order identity tensor on the space of symmetric second order tensors, the intact elastic modulus given by $\mathbb{C}$, the damaged elastic modulus by $\widetilde{\mathbb{C}}$, $\lambda >0 ,\mu >0$ the intact Lam\'{e} parameters, and $\widetilde{\lambda}$ and $\widetilde{\mu}$ the Lam\'{e} parameters for the damaged material, define
\[
\mathbb{C} := \lambda I \otimes I + 2\mu \mathbb{I}, \qquad \qquad \widetilde{\mathbb{C}} := \widetilde{\lambda}(|r|) I \otimes I + 2 \widetilde{\mu}(|r|) \mathbb{I}, \qquad \qquad \Delta \C_r := \C - \widetilde{\C},
\]
where $\widetilde{\lambda}, \widetilde{\mu}$ are positive, monotone decreasing functions of $|r|$ from the intact values of the parameters to some small (positive) residual values.
Let
\[
\widehat{r} := \frac{r}{|r|}, \qquad \qquad  \R := \widehat{r} \otimes \widehat{r} \otimes \widehat{r} \otimes \widehat{r}, \qquad \qquad E_r := \widehat{r} \cdot E \,\widehat{r},
\]
 $\rho_0 > 0$ be the mass density of the intact, unstretched elastic material, and $H(x) = 0$ for $x \leq 0$ and $H(x) = 1$ for $x > 0$ be the Heaviside function (an appropriately smoothed representation will also suffice). We now define the strain energy density of the material accounting for damage due to cracking as (cf. \cite{morin2021analysis})
\begin{equation}\label{eq:strn_energy}
\begin{aligned}
    2 \rho_0 \psi_e(W, c) & = H(|r|)\, E: \left[ \widetilde{\C}  + (1 - H(E_r)) \left( \Delta \C_r \cdot_4 \R \right) \R \right]: E  + \left(1 - H(|r|) \right) E : \mathbb{C} E\\
    & = H(|r|) \left[ E: \widetilde{\C} E + (1 - H(E_r))  (\Delta \C_r \cdot_4 \R) E_r^2 \right] + \left(1 - H(|r|) \right) E : \mathbb{C} E.
\end{aligned}
\end{equation}
$\psi_e$ has physical dimensions of energy per unit mass, and $A \cdot_4 B := A_{ijkl}B_{ijkl}$ for fourth-order tensors $A,B$.

It can be shown (using arguments given in \cite{acharya2011microcanonical} and \cite{morin2021analysis}) that for a crack-only model with $\psi = \psi_e(W,c) = \psi_e\left(F^TF,c \right)$ frame-indifferent, the `elastic-distortion driven' part of the Cauchy stress, say $T_{(F)}$, is given by $T_{(F)} = - \rho W^T \partial_W \psi_e = \rho F \partial_E \psi_e F^T$, and the normal stress component to the crack, $|c|^{-2} c \cdot T_{(F)} c$, at a damaged point where $H(|r|) = 1$ is given, up to a factor of $\frac{\rho |r|^2}{\rho_0 |c|^2} $, by $\widetilde{\lambda}(tr(E) - E_r) + \lambda E_r + 2 \mu E_r$ if the material point experiences compressive strain characterized by $H(E_r) = 0$, and by $\widetilde{\lambda}tr(E) + 2 \widetilde{\mu} E_r$ if the point experiences tensile normal strain perpendicular to the crack.

We also introduce a crack-resistance energy density function $\eta(|c|)$ with physical dimensions of energy per unit mass. A typical example reflecting no residual energy stored in damaged regions is
\begin{equation}\label{eq:eta_1}
\rho \, \eta(|c|) = \begin{cases}
a \left(1 - \cos^2\left(\pi \frac{|c|}{c_{sat}} \right) \right), & 0 \leq |c| \leq c_{sat} \\
0, & |c| \geq c_{sat},
\end{cases}
\end{equation}
where $a \geq 0$ (with physical dimensions of $stress$) and $c_{sat} > 0$ (dimensionless) are material constants. Another example, modeling Griffith type `surface energy' (but not dependent on crack-length) is
\begin{equation}\label{eq:eta_2}
\rho \, \eta(|c|) = \begin{cases}
a \left(1 - \cos^2\left(\pi \frac{|c|}{c_{sat}} \right) \right), & 0 \leq |c| \leq \frac{c_{sat}}{2} \\
a, & |c| \geq \frac{c_{sat}}{2}.
\end{cases}
\end{equation}
With these constructs a physically reasonable constitutive assumption for the free energy density of our material is
\begin{equation}\label{eq:free_energy}
\psi(W,\alpha, c, t, \rho) = \psi_e(W,c) + \eta(|c|) + \frac{\mu_\alpha \, l_\alpha^2}{2\rho} |\alpha|^2 + \frac{\mu_t \, l_t^2}{2\rho} |t|^2,
\end{equation}
where $0 < \mu_\alpha, \mu_t = \mathcal{O}(\mu)$ are material constants with dimensions of $stress$, and $l_\alpha, l_t > 0$ are material constants with dimensions of $length$. The lengths involved are expected to be much smaller than typical macroscopic dimensions.

Turning to the constitutive equation for the dislocation velocity in the bulk, motivated by the `driving force' for dislocation motion in \eqref{eq:sec_law_exp}, define
\[
\mathcal{PK} := X \left\{- \rho \, \dpsi{W} + \curl (\rho \, \dpsi{\alpha}) \right\}^T \alpha, \qquad \qquad p := \frac{X : (F\alpha)}{|X:(F\alpha)|},
\]
and a dislocation mobility tensor of the form
\begin{equation}\label{eq:disloc_mobility}
M = \frac{1}{|\alpha|^f}\left( m_{gl} (I - p \otimes p) + m_{cl} \, p \otimes p \right),
\end{equation}
where $f = 0$ or $1$, and $m_{gl}, m_{cl} \geq 0$ are material constants with physical dimensions of $\frac{length^2}{stress.time}$ for $f = 0$ and $\frac{length}{stress.time}$ for $f = 1$. Then we propose the constitutive assumption
\begin{equation}\label{eq:V_a}
\Va = M \mathcal{PK},
\end{equation}
and note that when $\psi$ is independent of $\alpha$, 
\[
\mathcal{PK} = X \left( T^T(F\alpha) \right),
\]
which is the natural generalization of the form of the Peach-Koehler force of classical dislocation theory to finite deformation.

For the crack velocity we assume a simple isotropic mobility:
\begin{equation}
\label{eq:V_t}
V^t = \frac{m_{cr}}{|t|^f}  \left\{ - \rho \dpsi{c} + \curl (\rho \dpsi{t}) \right\} \times t,
\end{equation}
where $m_{cr} > 0$ (with same physical dimensions as $m_{cl}$ or $m_{gl}$) is a material constant reflecting crack mobility and $f = 0$ or $1$. Ignoring the contribution of the second term in the crack driving force, We note that the crack velocity is not restricted to be in the direction of $t \times c$, allowing crack-tip motions off of the local crack-plane (defined by $c_\perp$).

Equations \eqref{eq:stress_reln}, \eqref{eq:free_energy}, \eqref{eq:V_a}, and \eqref{eq:V_t} form the constitutive assumptions of a specific model.

\section{Motivating questions for the development of the model}\label{sec:motiv}
Here, we outline some fundamental physical problems that served as the motivation for the development of the theory, and which can be used to evaluate its predictive capability through analysis and computation in the future.

\subsection{`Stokes flow' from nonlinear elasticity with defects}
Consider the quasi-static approximation for balance of linear momentum without body force:
\[
\divv T = 0.
\]
Since this holds for all times, it can be shown that this is equivalent to
\begin{equation}
\label{eq:quasi_static_rate}
    \divv \left[(\divv v) T + \dot{T} - TL^T \right] = 0,
\end{equation}
with $\divv T = 0$ initially. Assuming, for simplicity, that $\psi = \psi(F,c)$, we have that $T = T(F,c)$ so that $\dot{T} = \partial_F T : \dot{F} + \partial_c T \cdot \dot{c}$, and combining with \eqref{eq:W_evol} written in the form
\begin{equation}
  \label{eq:rate_decomp}  
  \dot{F}F^{-1} = L - (F\alpha) \times \Va - FL^p
\end{equation}
and \eqref{eq:c_evol} we obtain
\begin{equation}
    \label{eq:stress_rate}
    \dot{T} = \partial_F T: \left[ L - (F\alpha) \times \Va - FL^p \right]F + \partial_c T \cdot \left[ -L^T c + t \times V^t \right].
\end{equation}
Combining \eqref{eq:stress_rate} and \eqref{eq:quasi_static_rate} and defining
\begin{align*}
    \mathbb{L} & := T \otimes I - \mathbb{A} + \partial_F T F^T - \mathbb{B} \\
    \mathbb{A}_{ikrj} & = T_{ij} \delta_{kr}\\
    \mathbb{B}_{ijmk} & = \partial_{c_k} T_{ij} \, c_m,
\end{align*}
we have
\begin{equation}
    \label{eq:rate_form}
    \divv (\mathbb{L}:\nabla v) = \divv \left[ \partial_F T :\left\{ (F\alpha) \times \Va + FL^p \right\}F - \partial_c T \cdot (t \times V^t )  \right].
\end{equation}
Evidently, the evolution of cracks and dislocations play the role of a `body-force forcing' in the evolution of the deformation of the body.

In the presence of dislocations and cracks in general, but in the absence of their motion relative to the material, we have
\begin{equation}
    \label{eq:stokes}
    \divv(\mathbb{L}(F,c): \nabla v) = 0
\end{equation}
with standard combinations of Dirichlet b.c. on the velocity on the boundary and Neumann conditions related to the First Piola-Kirchhoff traction (w.r.t the current configuration as the reference) rate.

Evidently, \eqref{eq:stokes} does not reduce to an isotropic $4^{th}$-order tensor acting on the stretching tensor $L_{sym} = D$, but this system is energetically and mechanically (in terms of applied loads) reversible, whereas `viscous Stokes flow' is only mechanically reversible.

\subsection{Relation between defects in elastic solids and viscous fluids, and the mechanical load induced solid-fluid transition}
In an elastic solid, (dislocation) defects can be said to arise when the inverse elastic distortion is no longer curl-free, i.e.
\begin{equation*}
    -\curl W = \alpha \neq 0.
\end{equation*}
In a fluid one might say that defects arise when the velocity gradient develops a `singular part,' thinking, roughly, that the velocity field is discontinuous across 2-d surfaces.

What might be the connection between these two ideas? Can such a connection, in the context of a specific constitutive model, be used to study the transition of a solid to a fluid due to a proliferation of defects?

Noting \eqref{eq:W_evol} rewritten in the suggestive form of \eqref{eq:rate_decomp} and assuming $L^p = 0$ (a coarse-scale `homogenized' effect), when the $\alpha$ field is a distribution of superposed core fields moving with velocity $\Va$, $\alpha \times \Va$ very much looks like a singular distribution (when viewed macroscopically), and then \eqref{eq:rate_decomp} suggests that $\dot{F}F^{-1}$ - the elastic part of the velocity gradient (of the solid (fluid?)) - is its `regular part' (the absolutely continuous part), with $F(\alpha \times \Va)$ being its `singular' part.

\subsection{Dislocation nucleation}
Here we consider a model with no cracks and $L^p = 0$. Assume $\psi = \psi(F); T = T(F)$.
\subsubsection{Quasi-static balance of forces}\label{sec:qs_disloc_nucl}
The governing equations for $v, W, \alpha$ are:
\begin{subequations}\label{eq:qs_d_nucl}
\begin{align}
   \divv \left[ \left(T \otimes I - \mathbb{A} + \partial_F T F^T  \right): \nabla v \right] & = \divv \left[ \partial_F T: \left\{ (F\alpha) \times \Va \right\}\right] \label{eq:disloc_nucl_v}\\
   \dot{W} + WL & = \alpha \times \Va; \qquad \curl W = -\alpha \label{eq:disloc_nucl_W}\\
   \dot{\alpha} + (\divv v) \alpha - \alpha L^T & = - \curl \left( \alpha \times \Va \right) \label{eq:disloc_nucl_alpha}.
\end{align}
\end{subequations}
(although the fields $v, W$ suffice, intuition for nucleation related questions based on prior work suggests working with the $\alpha$ equation for this question).
\begin{itemize}
    \item \underline{Question}: Do perturbations in $\alpha$ from a dislocation-free state $\alpha = 0$ grow? Characterize the instability in terms of the class of elastic distortion fields $F$ and energy densities $\psi(F)$. The constitutive choices for $\Va$ can be as in \eqref{eq:V_a} and further simplified as necessary, e.g. assume isotropic mobility.
    
    The initial state satisfies $\divv T = 0$ and loading is required.
\end{itemize}
\subsubsection{Dynamic balance of forces}
In \eqref{eq:qs_d_nucl} replace \eqref{eq:disloc_nucl_v} with balance of linear momentum and balance of mass
\[
\divv T = \rho \dot{v}; \qquad \dot{\rho} + \rho \,\divv v = 0
\]
and ask the same question as in Sec. \ref{sec:qs_disloc_nucl}.

\subsection{Crack nucleation}
Here we consider a model with no dislocation or plasticity, $\alpha, L^p = 0$, and $\psi = \psi(F,c); T = T(F,c)$. Here, $W$ is a gradient on the current configuration and $F$ is as well, on the reference defined by the inverse deformation which is a potential for $W$ since $\curl W = 0$.
\subsubsection{Quasi-static balance of forces}\label{sec:qs_crack_nucl}
The governing equations for $v,W,c,t$ are:
\begin{subequations}\label{eq:qs_c_nucl}
\begin{align}
   \divv \left[ \left(T \otimes I - \mathbb{A} + \partial_F T F^T - \mathbb{B}  \right):\nabla v \right] & = - \divv \left[ \partial_c T \cdot\left( t \times V^t \right)\right] \label{eq:crack_nucl_v}\\
   \dot{c} + L^Tc & = t \times V^t \label{eq:crack_nucl_c}\\
   \dot{t} + (\divv v) t - Lt  & = - \curl \left( t \times V^t \right) \label{eq:crack_nucl_t}.
\end{align}
\end{subequations}
In the above $\dot{W} + WL = 0 \Rightarrow \dot{F}F^{-1} = L = \nabla v$. As in the dislocation case, one of \eqref{eq:crack_nucl_c} and \eqref{eq:crack_nucl_t} suffices, but can be used as necessary.
\begin{itemize}
    \item \underline{Question}: Do perturbations in $t$ from a crack-free state $ c = 0 \Longrightarrow t = 0$ grow? Characterize the instability in terms of the class of elastic distortion fields $F$ and energy densities $\psi(F)$. The constitutive choices for $V^t$ can be as in \eqref{eq:V_t}.
    
    The initial state satisfies $\divv T = 0$ and and loads are required.
\end{itemize}
\subsubsection{Dynamic balance of forces}
In \eqref{eq:qs_c_nucl} replace \eqref{eq:crack_nucl_v} with balance of linear momentum and balance of mass
\[
\divv T = \rho \dot{v}; \qquad \dot{\rho} + \rho \,\divv v = 0
\]
and ask the same question as in Sec. \ref{sec:qs_crack_nucl}.

\begin{remark}
For dislocation or crack nucleation from an undislocated or uncracked state, respectively, the corresponding evolution equations for the perturbation in dislocation density \textup{(}$\widetilde{\alpha}$\textup{)} and crack-tip density \textup{(}$\widetilde{t}$\textup{)} are given by
\[
\dot{\widetilde{\alpha}}_{i3} = - v_{r,r}\, \widetilde{\alpha}_{i3}, \qquad \dot{\widetilde{t}}_{3} = - v_{r,r}\, \widetilde{t}_{3}
\]
where, for simplicity, we assume that $i = 1,2$ and only straight dislocation/crack-tips in the 3-direction are allowed.

Based on the above, it seems that the distinction between crack and dislocation nucleation in this ansatz is a matter of nonlinear stability.  We note that for the purposes of linear stability, $\mathbb{B} = 0$ at the crack-free state.
\end{remark}

\subsection{Brittle-ductile transition}
\begin{figure}
\centering
\includegraphics[width=3.0in,height=3.5in]{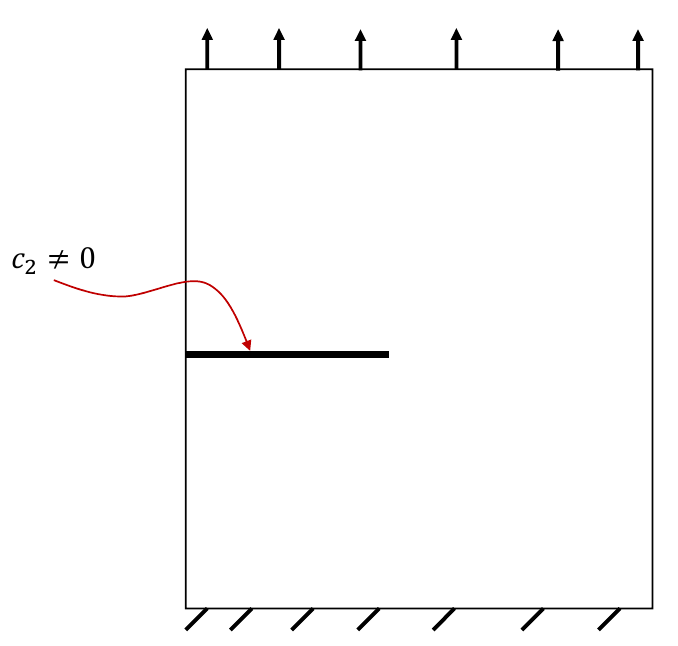}
\caption{Schematic of a body with an edge crack under load. In the brittle-ductile transition, the question is whether under load the crack propagates or a dislocation (dipole) nucleates and moves \cite{rice1974ductile}.}
\label{fig:edge_crack}
\end{figure}

This is a coupled crack-dislocation problem. The initial condition is that of an unloaded body with an edge crack as shown in Fig. \ref{fig:edge_crack}.
\begin{itemize}
    \item \underline{Question}: Under, say Mode I, loading, i.e., Dirichlet conditions on velocity $v_2 \neq 0$ on top boundary with bottom fixed as shown, does the stress field of the crack with a concentration at the notch-tip nucleate a dislocation (or a dipole) in the body which then moves (expands) causing plasticity (ductile behavior), or does the crack propagate without any dislocation nucleation and propagation (brittle behavior)? Characterize based on material parameters of the model (for a large body).
\end{itemize}

\subsection{Macroscopic model of elasto-viscoplasticity}\label{sec:evp}
Consider the dislocation-only model and define $\varepsilon := \frac{l_\alpha}{H}$, recall \eqref{eq:free_energy}, where $H$ is a representative dimension of the body and we will be interested in $\varepsilon \to 0$ with $l_\alpha$ \emph{fixed}.

Consider the system
\begin{align*}
0  &= \divv T^\varepsilon \\
\dot{W}^\varepsilon + W^\varepsilon L^\varepsilon & = - \curl W^\varepsilon \times (V^\alpha)^\varepsilon \\
& ``\Leftrightarrow" \\
\quad \dot{\alpha}^\varepsilon + tr(L^\varepsilon)\alpha^\varepsilon - \alpha^\varepsilon (L^\varepsilon)^T & = - \curl \left[\alpha^\varepsilon \times (V^\alpha)^\varepsilon\right]
\end{align*}
subjected to a constraint on the initial condition
\[
 \lim_{\varepsilon \to 0} \Vert \alpha_0^\varepsilon \Vert_{L^2(\Omega_0)} = \mbox{constant}_\alpha,
\]
a boundary condition
\begin{equation}\label{eq:v_bc}
   v^\varepsilon(\cdot, t) = \bar{v}(\cdot,t) \mbox{ on } \partial \mathcal{C} 
\end{equation}
where $\bar{v}$ is a given function, and
\[
\frac{1}{\mbox{vol} (\mathcal{C})} \int_{\mathcal{C}} \alpha^\varepsilon \, dv = 0 \quad \mbox{for all times} .
\]

\begin{itemize}
    \item \underline{Question}: (assuming existence of solutions for $\varepsilon > 0$, or plausible demonstration of such in finite-dimensional computational settings) What is the limit model that arises as $\varepsilon \to 0$? 

   Here, the `limit model' is the question of what model the evolution of the weak limits (roughly, all smoothly weighted space-time averages) of the fundamental fields of the model, say $\alpha^\varepsilon,v^\varepsilon$ here, satisfy. Since the microscopic equations are nonlinear and averages of nonlinear functions of field quantities ( say, e.g., $\alpha^\varepsilon \times V^\varepsilon$), do not equal the same functions evaluated on the averages of the said quantities, this requires the determination of the evolution of the weak limits of a further set of quantities, like say $|\alpha^\varepsilon|^2$, defined on a sequence of solutions of the microscopic model. Moreover, to be useful, such evolution of the limits must be `closed’ in the sense that it must need information only on the state of only the limits of these quantities at any given time.
        
    In particular, there is good physical intuition behind the expectation that $\lim_{\varepsilon \to 0} \alpha^\varepsilon \times (\Va)^\varepsilon$ produces an extra term in the limit, related to the plastic strain rate produced by the expansion of `sub-grid' loops, the latter not sensed by $\lim_{\varepsilon \to 0} \alpha^\varepsilon$. In fact, this is the reason for the phenomenological introduction of the term $L^p$ (and only this term as representative of the plastic strain rate) in macroscopic models.
    
    Is the limit parametrized by constant$_\alpha$?
\end{itemize}

\subsection{Macroscopic model of damage}
Consider the crack-only model and define $\varepsilon := \frac{l_t}{H}$, recall \eqref{eq:free_energy}, where $H$ is a representative dimension of the body and we will be interested in $\varepsilon \to 0$ with $l_t$ \emph{fixed}.

Consider the system
\begin{align*}
0  &= \divv T^\varepsilon \\
\dot{c}^\varepsilon + (L^\varepsilon)^T c^\varepsilon  & = - \curl c^\varepsilon \times (V^t)^\varepsilon \\
``& \Leftrightarrow" \\
\dot{t}^\varepsilon + tr(L^\varepsilon)t^\varepsilon - (L^\varepsilon)t^\varepsilon & = - \curl \left[t^\varepsilon \times (V^t)^\varepsilon\right].
\end{align*}

In the above, the equivalence is not strict since the bottom equation implies the top one up to the gradient of a scalar field which is assumed to vanish based on the assumption that microscopically the crack-tip flux can occur only in the presence of a crack-tip at a point (much like microscopic plastic strain rate/slipping rate at a point can arise only if a dislocation is present at a point).

Let the system be subjected to a constraint on the initial condition
\[
 \lim_{\varepsilon \to 0} \Vert c_0^\varepsilon \Vert_{L^2(\Omega_0)} = \mbox{constant}_c,
\]
a boundary condition
\[
v^\varepsilon(\cdot, t) = \bar{v}(\cdot,t) \mbox{ on } \partial \mathcal{C}
\]
where $\bar{v}$ is a given function, and
\[
\frac{1}{\mbox{vol} (\mathcal{C})} \int_{\mathcal{C}} t^\varepsilon \, dv = 0 \quad \mbox{for all times}.
\]

\begin{itemize}
    \item \underline{Question}: As in Sec. \ref{sec:evp}, what is the limit model as $\varepsilon \to 0$? Does a natural connection arise with the type of coupled brittle-ductile model of fracture proposed in \cite{acharya2020possible}?
\end{itemize}

\subsection{Classical elasto-viscoplasticity, viscoplasticity (a non-Newtonian viscous fluid), as limit models}
Consider the dislocation-only model from \eqref{eq:gov_eq}. The classical, phenomenological model of \textbf{\emph{elasto-viscoplasticity}} is given by the system \eqref{eq:gov_eq} with $\Va = 0$ and $L^p$ and $T(F)$ specified by constitutive assumptions. The strain-rate decomposition \eqref{eq:rate_decomp}, that follows from the conservation of Burgers vector \eqref{eq:a_evol}, then takes the form
\[
\dot{F}F^{-1} = L - FL^p.
\]
Recall that the model does not involve a reference configuration of any sort and $F$ is not a deformation gradient in general (customarily it is written as $F^e$, but a `multiplicative decomposition' of a deformation gradient from any reference plays no role in our development). Define
\begin{align*}
    (\dot{F}F^{-1})_{sym} =: D^e; & \qquad \qquad (\dot{F}F^{-1})_{skw} =: \omega^e\\
    L_{sym} =: D; & \qquad \qquad L_{skw} =: \omega\\
    (FL^p)_{sym} =: D^p; & \qquad \qquad (FL^p)_{skw} =: \omega^p
\end{align*}
The constitutive equation for $L^p$ specifies $D^p$ and $\omega^p$; in describing the elastoviscoplasticity of polycrystals without texture, it is customary to assume
\[
\omega^p = 0
\]
(but not for single crystals or strongly textured polycrystals).

Considering isotropic elasto-viscoplasticity for simplicity, a typical constitutive assumption for $D^p$ is
\begin{equation}
\label{eq:Dp}
    D^p = \frac{1}{\nu} \left( \frac{|T'|}{g}\right)^{\frac{1}{m}} \frac{T'}{|T'|},
\end{equation}
where $\nu$ has physical dimensions of $time$, $T'$ is the stress deviator, $m > 0$ is a dimensionless constant called the rate-sensitivity, and $g$, a scalar, is the \emph{strength} which may itself evolve; a common expression for metals is 
\begin{equation}\label{eq:voce}
   \dot{g} = \begin{cases} 
   \theta_0 \left( \frac{g_s - g}{g_s - g_0} \right)|D^p| & g < g_s \\
   0 & g = g_s,
   \end{cases}
\end{equation}
where $g_s \geq g_0 > 0, \theta_0 >0$ material constants, and $g_0$ is an initial value. For ice, the strength does not evolve, staying fixed at $g = g_0$. The rate-sensitivity, $m$, for ice is $\sim 4.0$, for metals usually small $\sim 0.01$.

One obtains the classical theory of \textbf{\emph{\textup{(}rigid\textup{)} viscoplasticity}} under the

\textbf{Assumption}: the elastic strain rate is `small', i.e.,
\[
\frac{\vert D^e \vert}{\vert D^p\vert} \ll 1,
\]
so that
\[
D = D^p
\]
is assumed. For the typical power-law constitutive behavior \eqref{eq:Dp}, one then has
\[
 \nu D =  \left( \frac{|T'|}{g}\right)^{\frac{1}{m}} \frac{T'}{|T'|} \Rightarrow \vert T' \vert = g (\nu |D|)^m; \qquad \qquad T' = g (\nu |D|)^m \frac{D}{|D|}
\]
which also implies incompressibility, and one obtains the constitutive behavior of a non-Newtonian viscous fluid
\[
T = -p I + g (\nu |D|)^m \frac{D}{|D|},
\]
where $p$ is the constitutively undetermined pressure.

\begin{itemize}
    \item \underline{Question}:  Can classical elasto-viscoplasticity and rigid-viscoplasticity, including the constitutive assumptions (\ref{eq:Dp}, \ref{eq:voce}), be recovered as particular limits of the model in Sec. \ref{sec:evp} and, if so, under what conditions? Presumably one such condition is $m_{cl} = 0$ in \eqref{eq:disloc_mobility}?
    \item \underline{Question}: Rate-independent behavior is assumed to arise in these models as $m \to 0$ in \eqref{eq:Dp}. Can this be justified as a limit when the rate of loading in \eqref{eq:v_bc} $\dot{\bar{v}} \to 0$? 
\end{itemize}

\section*{Acknowledgment}

This work was supported by the Center for Extreme Events in Structurally Evolving Materials, Army Research Laboratory Contract No. W911NF2320073, and NSF OIA-DMR grant \# 2021019. It is a pleasure to acknowledge discussions with Vladimir Sverak.
\section*{Appendix}
The argument used here may be called the `Ericksen equality' for the theory, extending an argument to dynamics of Ericksen \cite[sec.~VII]{ericksen1961conservation} in the context of continuum mechanics of nematic liquid crystals.

Consider a superposed rigid motion of the body on a given motion. For any pair of such motions, the value of the free energy density function at any material point remains unchanged at any arbitrarily chosen instant of time, $s$, i.e.,
\begin{equation}\label{eq:psi_inv}
\psi(x(X,s)) = \psi(R(s)x(X,s) + d(s)) \qquad \forall \mbox{ motions } x(\cdot \cdot,\cdot), \forall R(\cdot), d(\cdot).
\end{equation}
Under such a superposed rigid motion, the fields $W,\alpha,c,t, \rho$ transform as follows:
\[
W(s) \to W(s)R^T(s), \qquad \alpha(s) = \alpha(s)R^T(s), \qquad c(s) \to R(s) c(s), \qquad t(s) = R(s)t(s), \qquad \rho(s) \to \rho(s).
\]
These transformation rules are consistent with the evolution statements \eqref{eq:W_evol},\eqref{eq:c_evol},\eqref{eq:line_den},\eqref{eq:conservlaw}, if the field $\Va,V^t$ transform as objective vectors (and $L^p$ transforms as an objective 2-point tensor from the current configuration to a local elastic reference that is unaffected by rigid motions of the body).

Now consider the free energy density \eqref{eq:free_energy}, and arbitrarily fixed state $(W,\alpha,c,t, \rho)$ at an arbitrary instant of time $s$ and compute $\dot{\psi}(s)$ on a pair of rigidly associated motions as described above for which $R(s) = I$ and $\dot{R}(s) = S$, where $S$ is an arbitrarily fixed skew tensor, so that $\dot{R}(s) R^T(s) = S$. By \eqref{eq:psi_inv} the value of $\dot{\psi}(s)$ on both motions have to be equal which implies 
\begin{align}
& \left[\dpsi{W}:\dot{W} + \dpsi{\alpha}:\dot{\alpha} + \dpsi{c}:\dot{c} + \dpsi{t}:\dot{t} + \dpsi{\rho}:\dot{\rho} \right] = \dot{\psi} = \notag\\
 & \left[\dpsi{W}:\dot{W} + \dpsi{\alpha}:\dot{\alpha} + \dpsi{c}:\dot{c} + \dpsi{t}:\dot{t} + \dpsi{\rho}:\dot{\rho} \right] - \left[ \dpsi{W}:WS + \dpsi{\alpha}:\alpha S - \dpsi{c}\cdot S c - \dpsi{t} \cdot S t \right]  \notag\\
& \Rightarrow  \left[ W^T \dpsi{W} + \alpha^T \dpsi{\alpha} - \dpsi{c} \otimes c - \dpsi{t} \otimes t \right]:S = 0 \notag\\
& \Rightarrow \frac{1}{2} \left[ \left(W^T\dpsi{W} - \dpsi{W}^TW\right) + \left(\alpha^T \dpsi{\alpha} - \dpsi{\alpha}^T \alpha \right) - \left(\dpsi{c} \otimes c - c \otimes \dpsi{c} \right) - \left( \dpsi{t} \otimes t - t \otimes \dpsi{t} \right)\right] = 0. \notag
\end{align}
But this is exactly the skew part of the term highlighted in blue in \eqref{eq:sec_law_exp}.

\section*{Author Contribution}
All work was performed by the sole author.
\section*{Conflict of Interest}
I declare that I, the author, have no competing interests as defined by Springer, or other interests that might be perceived to influence the results and/or discussion reported in this paper.
\section*{Data Availability}
Not applicable. There is no relevant data generated for the paper.

\bibliographystyle{alpha}\bibliography{crack_plasticity}

\begin{thebibliography}{HADMC22}

\bibitem[AA20]{arora2020unification}
Rajat Arora and Amit Acharya.
\newblock A unification of finite deformation ${J}_2$ von-{M}ises plasticity
  and quantitative dislocation mechanics.
\newblock {\em Journal of the Mechanics and Physics of Solids}, 143:104050,
  2020.

\bibitem[Ach11]{acharya2011microcanonical}
Amit Acharya.
\newblock Microcanonical entropy and mesoscale dislocation mechanics and
  plasticity.
\newblock {\em Journal of Elasticity}, 104:23--44, 2011.

\bibitem[Ach18]{acharya2018fracture}
Amit Acharya.
\newblock Fracture and singularities of the mass-density gradient field.
\newblock {\em Journal of Elasticity}, 132(2):243--260, 2018.

\bibitem[Ach20]{acharya2020possible}
Amit Acharya.
\newblock A possible link between brittle and ductile failure by viewing
  fracture as a topological defect.
\newblock {\em Comptes Rendus. M{\'e}canique}, 348(4):275--284, 2020.

\bibitem[Asa83]{asaro1983micromechanics}
R.~J. Asaro.
\newblock Micromechanics of crystals and polycrystals.
\newblock {\em Advances in Applied Mechanics}, 23:1--115, 1983.

\bibitem[AZA20]{arora2020finite}
Rajat Arora, Xiaohan Zhang, and Amit Acharya.
\newblock Finite element approximation of finite deformation dislocation
  mechanics.
\newblock {\em Computer Methods in Applied Mechanics and Engineering},
  367:113076, 2020.

\bibitem[BC06]{bulatov2006computer}
Vasily Bulatov and Wei Cai.
\newblock {\em Computer simulations of dislocations}, volume~3.
\newblock OUP Oxford, 2006.

\bibitem[Eri61]{ericksen1961conservation}
J.~L. Ericksen.
\newblock Conservation laws for liquid crystals.
\newblock {\em Transactions of the Society of Rheology}, 5(1):23--34, 1961.

\bibitem[Fre98]{freund1998dynamic}
L.~B. Freund.
\newblock {\em Dynamic fracture mechanics}.
\newblock Cambridge university press, 1998.

\bibitem[GAM15]{garg2015study}
Akanksha Garg, Amit Acharya, and Craig~E. Maloney.
\newblock A study of conditions for dislocation nucleation in
  coarser-than-atomistic scale models.
\newblock {\em Journal of the Mechanics and Physics of Solids}, 75:76--92,
  2015.

\bibitem[HADMC22]{hakimzadeh2022phase}
Maryam Hakimzadeh, Vaibhav Agrawal, Kaushik Dayal, and Carlos Mora-Corral.
\newblock Phase-field finite deformation fracture with an effective energy for
  regularized crack face contact.
\newblock {\em Journal of the Mechanics and Physics of Solids}, 167:104994,
  2022.

\bibitem[Hav92]{havner1992finite}
K.~S. Havner.
\newblock {\em Finite plastic deformation of crystalline solids}.
\newblock Cambridge University Press, 1992.

\bibitem[HL82]{hirth_lothe}
J.~P. Hirth and J.~Lothe.
\newblock {\em Theory of dislocations}.
\newblock Krieger, 1982.

\bibitem[Hut79]{hutchinson_fracture}
John~W. Hutchinson.
\newblock {\em A course on nonlinear fracture mechanics}.
\newblock Dept. of Solid Mechanics, Technical University of Denmark, 1979.

\bibitem[MA21]{morin2021analysis}
L{\'e}o Morin and Amit Acharya.
\newblock Analysis of a model of field crack mechanics for brittle materials.
\newblock {\em Computer Methods in Applied Mechanics and Engineering},
  386:114061, 2021.

\bibitem[RT74]{rice1974ductile}
James~R. Rice and R.~Thomson.
\newblock Ductile versus brittle behaviour of crystals.
\newblock {\em The Philosophical Magazine: A Journal of Theoretical
  Experimental and Applied Physics}, 29(1):73--97, 1974.

\bibitem[SK19]{steinke2019phase}
C.~Steinke and M.~Kaliske.
\newblock A phase-field crack model based on directional stress decomposition.
\newblock {\em Computational Mechanics}, 63:1019--1046, 2019.

\bibitem[SSK22]{steinke2022energetically}
C.~Steinke, J.~Storm, and M.~Kaliske.
\newblock Energetically motivated crack orientation vector for phase-field
  fracture with a directional split.
\newblock {\em International Journal of Fracture}, 237(1-2):15--46, 2022.

\bibitem[ZAWB15]{zhang2015single}
Xiaohan Zhang, Amit Acharya, Noel~J. Walkington, and Jacobo Bielak.
\newblock A single theory for some quasi-static, supersonic, atomic, and
  tectonic scale applications of dislocations.
\newblock {\em Journal of the Mechanics and Physics of Solids}, 84:145--195,
  2015.

\end{thebibliography}

\end{document}